\documentstyle[prd,aps,epsf]{revtex}
\tighten
\begin{document}
\draft

\preprint{Imperial/TP/97-98/59}

\twocolumn[\hsize\textwidth\columnwidth\hsize\csname @twocolumnfalse\endcsname

\title{A Re-examination of Quenches in $^{4}$He (and $^{3}$He)}
\author{G.\ Karra and R.\ J.\ Rivers}
\address{Blackett Laboratory, Imperial College, London SW7 2BZ}
\date{\today}
\maketitle

\begin{abstract}
In the light of recent difficulties in observing vortices in
quenches  of liquid $^{4}He$ to its superfluid state we re-examine
the Zurek scenario for their production.  We argue that experiments
in $^{4}He$ are unlikely to produce true vortices in the numbers originally
anticipated, if at all, because of the wide
Ginzberg regime and the slowness of the mechanical quenches.
On the other hand, the observed production of unambiguous vortices in
neutron-bombarded $^{3}He$,
with its narrow Ginzberg regime, and fast quenches, is to be expected.
\end{abstract}


\vskip2pc]

In thermal equilibrium the behaviour of simple systems experiencing
a continuous phase transition is generic, as manifest in the utility
of Landau-Ginzburg theory.  Is this also true dynamically? The
relevance of this question is that the early universe proceeded
through a sequence of phase transitions whose consequences are
indirectly observable, but whose detailed dynamics is unknown. 
Although it is difficult to measure an order parameter as it
changes, many transitions generate topological charge or topological
defects which can be detected, and from which the evolution of the
field can be inferred.  Motivated in part by Kibble's
mechanism for the formation of cosmic strings in the early universe\cite{kibble1,kibble2},
Zurek suggested\cite{zurek1} that we measure the density of vortices produced
during quenches of liquid $^{4}He$ and $^{3}He$ into their superfluid
states.

The original scenario, as proposed by Zurek, is very simple. 
It is exemplified by assuming (as is roughly appropriate for $^{4}He$) 
that the dynamics
of the transition can be derived from an explicitly time-dependent
Landau-Ginzburg free energy of the form
\begin{equation}
F(t) = \int d^{3}x\,\,\bigg(\frac{-\hbar^{2}}{2m}|\nabla\phi |^{2}
+\alpha (t)|\phi |^{2} + \frac{1}{4}\beta |\phi |^{4}\bigg).
\label{F}
\end{equation}
In (\ref{F}) $\phi = (\phi_{1} + i\phi_{2})/\sqrt{2}$ is the
complex order-parameter field, whose magnitude determines the
superfluid density.   In equilibrium at temperature
$T$, in a mean field approximation,  the chemical potential $\alpha (T)$ takes the form
$\alpha (T) = \alpha_{0}\epsilon (T_{c})$, where $\epsilon = (T/T_{c}
-1)$ measures the critical temperature
$T_{c}$ relative to $T$. 
In a quench in which $T_{c}$ or $T$ changes we
assume that $\epsilon$ can be written as
\[
\epsilon (t) = \epsilon_{0} - \frac{t}{\tau_{Q}}\theta (t)
\]
for $-\infty < t < \tau_{Q}(1 + \epsilon_{0})$, after which
$\epsilon (t) = -1$.  $\epsilon_{0}$  
measures the original relative temperature and $\tau_{Q}$
defines the quench rate.  The quench begins at time $t = 0$ and the
transition from the normal to the superfluid phase begins at time $t
= \epsilon_{0}\tau_{Q}$.

With $\xi_{0}^{2} = \hbar^{2}/2m\alpha_{0}$ and $\tau_{0} = \hbar
/\alpha_{0}$ setting the fundamental distance and time scales, the
equilibrium correlation length $\xi_{eq} (\Delta t)$ and the relaxation
time $\tau (\Delta t)$ diverge when the relative time $\Delta t = t
-\epsilon_{0}\tau_{Q}$ vanishes as 
\[
\xi_{eq} (\Delta t) = \xi_{0}\bigg|\frac{\Delta t}{\tau_{Q}}\bigg|^{-1/2},
\,\,\tau(\Delta t) = \tau_{0}\bigg|\frac{\Delta t}{\tau_{Q}}\bigg|^{-1}.
\]
Meanwhile, the non-equilibrium correlation length $\xi (\Delta t)$ at relative time
$\Delta t$ is defined from the diagonal {\it equal-time} correlation
function (up to powers)
\[
\langle \phi_{a}({\bf r})\phi_{b}({\bf 0})\rangle_{\Delta t} =
\delta_{ab}G(r,\Delta t)\sim \delta_{ab}e^{-r/\xi (\Delta t)},
\]
for $r$ large compared to $\xi$ (but not necessarily
asymptotically large).  As we approach the transition, eventually the relaxation time will
be so long that the system will not be able to keep up with the
pressure or temperature change.  We estimate the  relative
time $-{\bar t}  < 0$ at which the change from
equilibrium to non-equilibrium behaviour occurs by identifying
$\tau(-{\bar t})$ with ${\bar t}$ {\it i.e.}
${\bar t} = \sqrt{\tau_{0}\tau_{Q}}$.  After this time it is
assumed that the relaxation time is so long  that
$\xi (\Delta t)$ is more or less frozen in at the value
$\xi_{-}:=\xi (-{\bar t})\approx \xi_{eq} (-{\bar t}) =
\xi_{0}(\tau_{Q}/\tau_{0})^{1/4}$  until
the system is again changing slowly, at time $\Delta t \approx
+{\bar t}$, when it attains the value $\xi_{+}:=\xi (+{\bar t})\approx\xi_{eq} (+{\bar
t})=\xi_{eq} (-{\bar t})$.

If, for the sake of argument, we assume that the field fluctuations are
approximately Gaussian, then the  field
phases are correlated as
$\langle e^{i\theta({\bf r})}\,e^{-i\theta({\bf 0})}\rangle_{\Delta t} \sim
e^{-r/\xi (\Delta t)}$, where
 $\phi_{a}({\bf r}) = |\phi_{a}({\bf
r})|\,e^{i\theta ({\bf r})}$, on the
same scale as the fields.
The correlation length of the field can only be measured indirectly,
but the field fluctuations will encode vortices as defects in the phase. If the {\it initial}
vortex separation (in a plane) is $\xi_{def}$, then their density, $n_{def}$, is
$n_{def} = O(1/\xi_{def}^{2})$.
Zurek initially made the assumption that 
$\xi_{def}\approx\xi_{+}$ whereby, 
with $\xi_{+}\approx\xi_{-}$ in turn,
\begin{equation}
n_{def} = 
O\bigg(\frac{1}{\xi_{0}^{2}}\sqrt{\frac{\tau_{Q}}{\tau_{0}}}\bigg).
\label{nZ}
\end{equation}
Since $\xi_{0}$ also measures cold vortex thickness, $\tau_{Q}\gg
\tau_{0}$ corresponds to a measurably large number of widely
separated vortices.  We note that, since the argument is basically
one of causality, a free
energy of the form (\ref{F}) is not strictly required.  

The simple order of magnitude prediction (\ref{nZ}), with RG
improvement for $^{4}He$, has been the motive for several experiments. In
vortex production in quenches of $^{3}He$
into its $B$-phase two separate experiments at
Grenoble\cite{grenoble} and Helsinki\cite{helsinki} show quantitative
agreement with (\ref{nZ}) and one
experiment\cite{lancaster} at Lancaster on vortex production in
$^{4}He$ shows consistency.  However, a second 
experiment\cite{lancaster2} on $^{4}He$ finds no vortices.  
Because of this, it is timely to reexamine this
picture, with its implicit assumption of an initial domain structure
characterised by a single length.
In this brief note we shall argue that the prediction (\ref{nZ})
is unreliable because vortex separation cannot be related simply to
field or phase correlation length.  However, the success of the
$^{3}He$ experiments shows that there can be a {\it de facto}
similarity, and we shall show how this arises.

To see this we need an explicit
model, and again we adopt (\ref{F}).
Motivated by Zurek's later numerical\cite{zurek2} simulations with the time-dependent
Landau-Ginzburg (TDLG) equation for $F$ of (\ref{F}), 
we assume a linear response
\begin{equation}
\frac{1}{\Gamma}\frac{\partial\phi_{a}}{\partial t} = -\frac{\delta
F}{\delta\phi_{a}} + \eta_{a},
\label{tdlg}
\end{equation}
where $\eta_{a}$ is Gaussian noise. We go further and assume that, for early times, prior
to the completion of the symmetry breaking at least,
the self-interaction term can be neglected ($\beta =0$).  This both
preserves Gaussian field fluctuations and leads to $\xi_{\pm}\approx\xi_{0}(\tau_{Q}/\tau_{0})^{1/4}$ arising
in a natural way, as we shall see.  

Firstly, we
anticipate that correlation lengths are
approximately frozen
in during the time interval $-{\bar t}\leq \Delta t\leq {\bar t}$. 
Thus,  in the
first instance it is sufficient to perform our calculations at  $\Delta t =
0$, for which $D:=\xi(\Delta t = 0)\approx\xi_{+}\approx\xi_{-}$.  Writing the resultant equation
in time and space units $\tau_{0}$ and $\xi_{0}$ as
\[
{\dot \phi}_{a}({\bf k},t) = -\bigg[k^{2} + \bigg({\epsilon_{0} -
\frac{t}{\tau_{Q}}\theta (t)\bigg)\bigg]\phi_{a}(\bf k},t)
+\tau_{0}\eta_{a}({\bf k},t) 
\]
gives the solution, at $\Delta t =0$, $\phi_{a}({\bf k})=$
\[
 =\tau_{0}\int_{-\infty}^{\epsilon_{0}\tau_{Q}}dt
\,\exp\bigg[-\int_{t}^{\epsilon_{0}\tau_{Q}}dt'\bigg[k^{2} + \bigg(\epsilon_{0} -
\frac{t'}{\tau_{Q}}\theta (t)\bigg)\bigg]\eta_{a}({\bf k}).
\]
The resulting un-normalised correlation function has power 
(Fourier transform)
\begin{equation}
G(k) = \int_{-\infty}^{\epsilon_{0}\tau_{Q}}dt
\exp\bigg[-2\int_{t}^{\epsilon_{0}\tau_{Q}}dt'\bigg[k^{2} + \bigg(\epsilon_{0} -
\frac{t'}{\tau_{Q}}\theta (t')\bigg)\bigg].
\label{Gk}
\end{equation}
For a typical quench in $^{4}He$, $\epsilon_{0}\sim 10^{-2}$ 
is very small, but $\tau_{Q}\sim 10^{10}$ is so large that
$\epsilon_{0}\tau_{Q},\,\,\epsilon_{0}^{2}\tau_{Q}\gg 1$. 
$G(k)$ is then approximately independent of $\epsilon_{0}$, 
\[
G(k)=e^{\tau_{Q}k^{4}}\int_{\tau_{Q}k^{2}}^{\tau_{Q}
(\epsilon_{0} + k^{2})}dt\,e^{-t^{2}/\tau_{Q}}.
\]
This does not lead to an {\it asymptotic} fall-off of
the form $e^{-r/D}$. In fact, for large $r$, $G(r)\propto
\exp(-O((r/D)^{4/3}))$.  
Nonetheless, numerically, it is remarkably well
represented by $e^{-r/D}$, with coefficient unity in the exponent,
for $r$ being a few multiples of
$D$, for reasons that are not clear to us.  In that sense
Zurek's prediction for a correlation length of the form 
$D = \xi_{0}(\tau_{Q}/\tau_{0})^{1/4}$ is robust,
since explicit calculation\cite{ray2} shows that 
 $\xi (t)$ does not vary strongly in the interval
$-{\bar t}\leq \Delta t\leq{\bar t}$.

Despite this approximate constancy, 
the relationship between the unobservable ${\xi (t)}$ and the observable $\xi_{def}$
is, a priori, very weak since the two measure very different
attributes of the field fluctuations.  
Whereas the core of every vortex is a line zero of the complex
field $\phi$ the converse is not true since zeroes occur on
all scales.  However, a starting-point for counting
vortices in superfluids
is to count
line zeroes of an appropriately coarse-grained field, in which
structure on a scale smaller than $\xi_{0}$, the classical vortex size, is
not present\cite{popov}.
This is, indeed, the
basis of the numerous numerical simulations\cite{tanmay} of vortex
networks built from Gaussian fluctuations (but see \cite{gleiser}). 
For the moment, we put in a cutoff $l = O(\xi_{0})$ by hand, as
\[
G(r) = \int d \! \! \! / ^3 k\, e^{i{\bf k}.{\bf x}}G(k)\,e^{-k^{2}l^{2}}.
\]
We stress that  the {\it long-distance} correlation length
$D$  depends only on the position of the
nearest singularity of $G(k)$ in the complex k-plane, {\it
independent} of $l$. 

This is not the case for the line-zero density $n_{zero}$, 
depending, in our Gaussian approximation\cite{halperin,maz}, on
the {\it short-distance} behaviour of $G(r)$,
\begin{equation}
n_{zero} = \frac{1}{2\pi\xi_{zero}^{2}} = \frac{-1}{2\pi}\frac{G''(0)}{G(0)},
\label{ndef}
\end{equation}
the ratio of fourth to second moments of $G(k)\,e^{-k^{2}l^{2}}$. 
In evaluating
\[
\xi_{zero}^{2} = \int_{0}^{\infty}dk\,k^{2}\,e^{-k^{2}l^{2}}G(k)\bigg/
\int_{0}^{\infty}dk\,k^{4}\,e^{-k^{2}l^{2}}G(k)
\]  
we substitute for $G(k)$ from (\ref{Gk}) and perform the $k$
integration first.  For small $\epsilon_{0}$, very large $\tau_{Q}$,
and $l = O(1)$, in dimensionless units the dominant contribution is
from $t\approx\epsilon_{0}\tau_{Q}$. On neglecting terms relatively 
$O(e^{-\epsilon_{0}^{2}\tau_{Q}})$ we find, at $\Delta t =0$,
\begin{equation}
\xi_{zero}^{2}\propto
\int_{0}^{\infty}dt\frac{e^{-t^{2}/\tau_{Q}}}{[t +l^{2}/2]^{3/2}}\bigg/
\int_{0}^{\infty}dt\frac{e^{-t^{2}/\tau_{Q}}}{[t
+l^{2}/2]^{5/2}}\approx O(l^{2}),
\label{xidef0}
\end{equation} 
independent of $\epsilon_{0}$.
The details are immaterial.  Firstly, since $\xi_{zero}\ll D$ the
frozen correlation length $D$ of the field, and its phase, 
does not set the scale at which line zeroes appear.  
Equally importantly, we have a situation in which the
density of line zeroes depends entirely on the scale at which we look. 
Such fractal behaviour cannot be
understood as representing vortices. We would not wish to
identify the length scale $\xi_{zero}$ with
$\xi_{def}$ at this time, even if the latter could be defined, or $D$ with either.  

This is not surprising.  Although the field correlation length
$\xi (\Delta t)$ may
have frozen in at $D$ by $\Delta t = 0$, the symmetry breaking has not been effected. 
In a classical sense at least, vortices can only be identified once
the field magnitude has grown to its equilibrium value
\begin{equation}
\langle |\phi |^{2}\rangle = \alpha_{0}/\beta,
\label{eq}
\end{equation} 
and we should not begin to count them before then.  Even though the
long-range correlation length $\xi (\Delta t)$ will not change
substantially in that time, this will not be the case for $\xi_{zero}(\Delta
t)$, as long-range modes increase in amplitude as the field becomes ordered.

Specifically, on continuing to use (\ref{tdlg}) for times $t >
\epsilon_{0}\tau_{Q}$ we see that,
as the unfreezing occurs, long wavelength modes with $k^{2} < t/\tau_Q -
\epsilon_{0}$ grow exponentially.  Provided
$\epsilon_{0}^{2}\tau_{Q}\gg 1$ they soon begin to dominate the
correlation functions.  Let 
\[
G_{n}(\Delta t ) = \int_{0}^{\infty}dk\,k^{2n}\,G(k,\Delta t)
\]
be the moments of
$G(k,\Delta t)$,
now of the form $G(k,\Delta t)=$ 
\[
\int_{-\infty}^{\epsilon_{0}\tau_{Q} +\Delta t}dt'
\exp\bigg[-2\int_{t'}^{\epsilon_{0}\tau_{Q} +\Delta t }dt''\bigg[k^{2} + \bigg(\epsilon_{0} -
\frac{t''}{\tau_{Q}}\theta (t'')\bigg)\bigg].
\]
We find that
\begin{equation}
G_{n}(\Delta t)\approx\frac{I_{n}}{2^{n + 1/2}}\,e^{(\Delta t/{\bar t})^{2}}
\int_{0}^{\infty}dt'\,\frac{e^{-(t'-\Delta t)^{2}/{\bar t}^{2}}}{[t'
+l^{2}/2]^{n +1/2}},
\label{Gt}
\end{equation}
where we measure the dimensionless time $\Delta t$ in units of ${\bar t} = \sqrt{\tau_{Q}}$
from $t = \epsilon_{0}\tau_{Q}$ and $I_{n} = \int_{0}dk k^{2n}\, e^{-k^{2}}$.
For small relative times the integrand gets a large
contribution from the ultraviolet cutoff dependent lower endpoint, and we recover
(\ref{xidef0}).
As long as the endpoints make a significant contribution to the whole
then the density of line zeroes derived from (\ref{Gt}) will be
strongly dependent on scale.
Only when their contribution is small and $\partial n_{zero}/\partial l$ is
small in comparison to
$n_{zero}/l$ at $l = \xi_{0}$ can we identify the essentially
non-fractal line-zeroes
with vortices, and $\xi_{zero}$ with $\xi_{def}$. 

$\Delta t_{1}$ and $\Delta t_{2}$,  the times at which the exponential modes begin
to dominate in the integrands of $G_{1}$ and $G_{2}$, can be determined 
by comparing the relative strengths of the contributions from the
scale-independent saddlepoint and the endpoint in (\ref{Gt}). 
If $p_{1}= \Delta t_{1}/{\bar t}$ and $p_{2}= \Delta t_{2}/{\bar t}$
are the multiples of ${\bar t}$ at which it happens then,
for the former to
dominate the latter requires 
$e^{p_{1}^{2}}/p_{1}^{3/2}> \tau_{Q}^{1/4}/\sqrt{2\pi}$ and
$e^{p_{2}^{2}}/p_{2}^{5/2}> 3\sqrt{2}\tau_{Q}^{3/4}/\sqrt{\pi}$.
We see that it takes longer for the long wavelength modes to
dominate $G_{2}$ than the order parameter $G_{1}$.
Because of the exponential growth, $p_{1}$ and $p_{2}$ are $O(1)$.
We stress that these are
lower bounds.

Since $\xi (\Delta t)$ and $\xi_{def}(\Delta t)$ 
use different attributes of the power spectrum
$G(k)\,e^{-k^{2}l^{2}}$ there is no {\it a priori} reason as to why they
should be related. However,
if the linear equation (\ref{Gt}) were valid for later times $\Delta
t>\Delta t_{2}$  then the integrals are dominated by the
saddle-point at $t'=\Delta t$, to give a separation of line
zeroes $\xi_{zero}(\Delta t )$ of the form
\begin{equation}
\xi_{zero}^{2}(\Delta t) = \frac{G_{1}(\Delta t)}{G_{2}(\Delta t)}
\approx\frac{4\Delta t}{3{\bar t}}\xi_{0}^{2}\bigg(\frac{\tau_{Q}}{\tau_{0}}\bigg)^{1/2}
=\frac{4\Delta t}{3{\bar t}}D^{2}
\label{newZ}
\end{equation}
approximately {\it independent} of the cutoff $l$ for $l = O(\xi_{0})$.
Only then, because of the transfer of
power to long wavelengths, do  line zeroes become widely separated, and
$\xi_{zero}(\Delta t)\approx\xi_{def}(\Delta t) $ does begin to measure vortices. 
Thus, if the order parameter is large enough that it takes a long
time (in units of ${\bar t}$) for the field to populate its
ground states (\ref{eq}) we would recover the result
$\xi_{def}= O(D)$ as an order of magnitude result from (\ref{newZ}), but
for different reasons. 

Whether we have time enough depends on the self-coupling $\beta$,
which determines when the linear approximation fails.
At the absolute latest, the correlation function
must stop its exponential growth at $\Delta t = \Delta t_{sp}$, when $\langle |\phi |^{2}\rangle$,
proportional to $G_{1}$, satisfies (\ref{eq}).  Let us suppose that
the effect of the backreaction that stops the growth initially
freezes in any defects.  This then is our prospective starting point for identifying
and counting vortices.  

To determine $\Delta t_{sp}$ we need to know the strength of the noise.  In
our dimensionless units we have
\[
\langle\eta_{a}({\bf k}, t)\eta_{b}(-{\bf k}', t')\rangle
= \delta_{ab}(2k_{B}T)(\xi_{0}^{3}/\alpha_{0})\delta (t-t')
\delta \! \! \! / ^3 ({\bf k} - {\bf k}').
\]
As a result, (\ref{eq}) is satisfied when
\begin{equation}
G_{1} (\Delta t_{sp})= \pi^{2}\alpha_{0}^{2}\xi_{0}^{3}/\beta k_{B}T_{c}
= \pi^{2}/\sqrt{1-T_{G}/T_{c}},
\label{Gmax}
\end{equation}
and we have used the fact, for $^{4}He$ at least, that the absolute temperature does not
change dramatically during the quench.
$T_{G}$ is the Ginzburg temperature, above which the potential 
difference between the false vacuum $\phi = 0$ and the
symmetry-broken ground states is sufficiently small to be easily
bridged by thermal fluctuations in a correlation volume.  
 
In order that $G_{1}$ is dominated by the exponentially growing
modes, so as to recover the modification of prediction (\ref{nZ}) 
in the form (\ref{newZ}) for identifiable vortices, the condition $\Delta t_{sp}>\Delta t_{1},
\Delta t_{2}$ becomes, from (\ref{Gmax})
\begin{equation}
(\tau_{Q}/\tau_{0})(1-T_{G}/T_{c})<C\pi^{4},
\label{tsp}
\end{equation}
on restoring $\tau_{0}$, where $C\approx 1$.
How strongly the inequality should be satisfied in (\ref{tsp}) is not obvious,
assuming as it does that the backreaction is effectively
instantaneous, and that mean field critical indices are valid. 
The former  probably means that the density above is an
overestimate if self-consistent linearity works here as in the
relativistic theory\cite{ray}. On the other hand, 
the interaction between field modes will tend to
redistribute the power back to shorter wavelengths in the
short-term, and the retarded nature of $G(r)$ after $\Delta t_{sp}$
will impose oscillations that may have consequences, as well as the
possible incorporation of phonon modes.
However, there is no way that the inequality can be remotely
satisfied for $^{4}He$, when subjected to a slow mechanical quench, as in
the Lancaster experiment, for which $\tau_{Q}/\tau_{0} =
O(10^{10})$, since the Ginzburg regime is so large 
that $(1-T_G /T_c ) = O(1)$.  
That is, field growth must stop long before  vortices are  well
defined.  Further, since the experiment leaves the superfluid in the
Ginzburg regime, thermal fluctuations will inhibit the creation of stable defects,
although it may be that
incoherent fluctuations, even if not vortices, will give a signal. [In no
experimemt is the vortex separation given by the corresponding
Ginzburg correlation length $\xi_{G} = \xi_{0}/\sqrt{1-T_{G}/T_{c}}$,
independent of $\tau_{Q}$, as was originally thought\cite{kibble1}].  
Adopting renormalisation improved critical indices cannot repair
such a deficit.
Whatever, there is no
reason to expect a vortex density (\ref{nZ}).

The situation for $^{3}He-B$ is potentially very different,  
with a much more complicated $SO(3)\times SO(3)\times U(1)$ order
parameter $A_{ai}$\cite{volovik} permitting several types of vortex, whose
cores are not necessarily normal.  However, for rapid quenches a 
 TDLG approach is valid\cite{Bunkov} with potential $V(A) = \alpha
(t)|A_{ai}|^{2} + O(A^{4})$. Prior to back-reaction
setting in, this is similar to the $^{4}He$ case
considered above in (\ref{F}), although self-consistent linearity
would be somewhat different. Suppose that (\ref{tsp}), which looks like an inequality between
the quench rate and the equilibriation rate that permits unstable
modes time to dominate over short-range fluctuations, remains true, albeit
with new coefficients.
Firstly, for $^{3}He-B$, 
the Ginzburg regime is very small, with $1-T_{G}/T_{c} =
O(10^{-8})$ and mean-field is a good approximation.
Secondly, in generating the phase transition by nuclear reactions\cite{grenoble,helsinki},
rather than by mechanical expansion, 
the quench rate is increased dramatically, with 
$\tau_{Q} = 10^{2} - 10^{4}$.
The inequality (\ref{tsp}) is satisfied by a huge margin and we can understand the success of the Helsinki and
Grenoble experiments.
We note that, if the density is given by (\ref{newZ}), then this 
overestimates the primordial density (\ref{nZ}) as
\begin{equation}
n_{def}\approx 
\frac{3}{8\pi p}\frac{1}{\xi_{0}^{2}}\sqrt{\frac{\tau_{Q}}{\tau_{0}}},
\label{nZ2}
\end{equation}
where $p =\Delta t_{sp}/{\bar t} > 1 = O(1)$.

Much more detailed numerical modelling in {\it three} spatial dimensions is needed before we can draw
any quantitative conclusions at later times. In $d<3$ dimensions
$G_{1}$ and $G_{2}$ are replaced by $G_{(d-1)/2}$ and $G_{(d+1)/2}$.
It is
easier to create defects in less than three dimensions
(particularly in one dimension\cite{dziarmaga}) since, with smaller $n$, the
saddle point in (\ref{Gt}) dominates more easily over the weakened ultraviolet endpoint.  

However, we believe that we have presented a convincing case that,
when line-zero density is well-defined, it can be equated to vortex
density.  Because line-zero density is based on
different attributes of the power $G(k)$ from the effective
long-range correlation length $\xi (\Delta t)$, we should not equate
them directly. Only if there is time enough before
the field populates the ground states do we recover the 
result (\ref{nZ2}), a specific case of (\ref{nZ}), although for different reasons. Moreover, the necessary
condition (\ref{tsp}) discriminates cleanly between the experiments
on $^{4}He$ and $^{3}He$. More details will be given
elsewhere\cite{ray2}.

Finally, a constraint like (\ref{tsp}) is not specific to the TDLG
theory.  A similar approach to weakly coupled
relativistic quantum
field theory\cite{ray} permits a comparison with Kibble's
predictions\cite{kibble2}.  After some rearrangement, it can be seen
\cite{ray} that a necessary condition for
the appearance of well-defined $O(2)$ vortices after the immediate
implementation of the transition is that
$(\tau_{Q}/\tau_{0})^{2}(1-T_{G}/T_{c})<1$, although we did not
appreciate this at the time.

We thank Alasdair Gill, Tom Kibble, Wojciech Zurek and Yuriy Bunkov for fruitful discussions.
G.K. would like to thank the Greek State Scholarship Foundation (I.K.Y.) for
financial support. This work is the result of a network supported by the European
Science Foundation.

\end{document}